\documentclass[pre,amsmath]{revtex4-2}
\usepackage{graphicx}
\usepackage{color}
\usepackage{bm}
\usepackage{epsfig}
\usepackage{latexsym}
\usepackage{nameref,hyperref}
\usepackage{pifont}
\usepackage[utf8]{inputenc}
\usepackage{subcaption}
\usepackage{csquotes}
\usepackage{enumitem}
\usepackage{bm}
\usepackage{float}
\usepackage{xcolor}
\usepackage{multirow}
\usepackage{amsmath}

\begin{document}
\title{Effect of relative timescale on a system of particles sliding on a fluctuating energy landscape: Exact derivation of product measure condition}
\author{Chandradip Khamrai and Sakuntala Chatterjee}
\affiliation{Department of Physics of Complex Systems, S. N. Bose National Centre for Basic Sciences, Block JD, Sector 3, Salt Lake, Kolkata 700106, India.}

\begin{abstract}

We consider a system of hardcore particles advected by a fluctuating potential energy landscape, whose dynamics is in turn affected by the particles. Earlier studies have shown that as a result of two-way coupling between the landscape and the particles, the system shows an interesting phase diagram as the coupling
parameters are varied. The phase diagram consists of various different kinds of ordered phases and a disordered phase. We introduce a relative timescale $\omega$ between the particle and landscape dynamics, and study its effect on the steady state properties. We find there exists a critical value $\omega = \omega_{c}$ when all configurations of the system are equally likely in the steady state. We prove this result exactly in a discrete lattice system and obtain an exact expression for $\omega_c$ in terms of the coupling parameters of the system. We show that $\omega_c$ is finite in the disordered phase, diverges at the boundary between the ordered and disordered phase, and is undefined in the ordered phase. We also derive $\omega_c$ from a coarse-grained level description of the system using linear hydrodynamics. We start with the assumption that there is a specific value $\omega^\ast$ of the relative timescale when correlations in the system vanish, and mean-field theory gives exact expressions for the current Jacobian matrix $A$ and compressibility matrix $K$. Our exact calculations show that Onsager-type current symmetry relation $AK = KA^{T}$ can be satisfied if and only if $\omega^\ast = \omega_c$ . Our coarse-grained model calculations can be easily generalized to other coupled systems.

\end{abstract}
\maketitle

\section{Introduction}

Models of coupled driven systems are often used to study several complex phenomena in physics and biology \cite{gnesotto2018broken, sediments, fang2019nonequilibrium}. Examples include particles advected by turbulent fluid \cite{parsa2012rotation, elperin1998dynamics, chatterjee2006dynamics, fluctuationdominatedphase}, shape deformation of cell membrane by curvature sensitive membrane proteins \cite{gov2006dynamics, veksler2007phase, veksler2009calcium, kabaso2011theoretical, peleg2011propagating, fovsnarivc2019theoretical, cagnetta2018active}, sedimenting colloidal crystal through a dissipative medium \cite{sediments, lahiri1997steadily}, etc. Also, recently therehas been a surge of interest in coupled nonequilibrium systems with multiple conserved quantities \cite{fibonacci, popkovuniversality, popkovexact, saito2021microscopic} since it was shown using the tools of nonlinear fluctuating hydrodynamics \cite{spohn2014nonlinear, mendl2013dynamic, van2012exact, fibonacci} that such systems often show unusual dynamical universality classes, apart from the well-known diffusive and Kardar-Parisi-Zhang (KPZ) class. In particular, it was shown that the dynamical exponent which characterizes the universality class can be expressed as the Kepler ratio of successive numbers of the Fibonacci sequence \cite{fibonacci}. Although most of these models follow rather simple dynamical rules, predicting the state of the system in the long-time limit is often a challenging task.

In this paper, we consider a lattice model of “light” and “heavy” particles moving on a fluctuating energy landscape in one dimension. The heavy (light) particles prefer to move towards the local energy minima (maxima), and the shape of the local landscape gets modified by the presence of these heavy and light particles. This model is known as the light-heavy model or LH model, and was introduced in \cite{chakrabortylarge}. It was shown that by varying the nature of coupling between the particle motion and the landscape dynamics, a rich phase diagram can be obtained, consisting of various kinds of ordered phases and a disordered phase \cite{chakrabortylarge}. The static and dynamic properties of these phases were characterized in \cite{static, dynamic, chakrabortyUniversality}. While the ordered phases show long-range order among the particles and in the landscape, in the disordered phase only short-range correlations exist in the system. However, the exact form of these short-range correlations is not known in general. Often these correlations are quite strong, because of which simple analytical approximations such as mean-field theory fail. In the absence of any exact knowledge of the nonequilibrium steady state, and breaking down of mean-field approximations, it becomes almost impossible to perform any exact, or even approximate analytical calculation in the presence of these short-range correlations.

In an attempt to understand these short-ranged correlations better, in this work we introduce a relative time-scale in the coupled dynamics of particle slide and landscape movement. This means compared to the landscape movement, the particle movement now happens faster, or slower. If the particles move much faster than the landscape, then the particles get enough time to find their preferred locations on the landscape. This is not possible in the opposite limit when the landscape movement happens faster. Therefore, the choice of the relative time-scale has a strong effect on the nature of the steady state \cite{sadhu2016actin, sadhu2018actin, passiveslider1, passiveslider2, singleactiveslider} and hence on the short-ranged correlations described above. We investigate this effect in this work, using exact calculations and Monte Carlo simulations.

In our Monte Carlo simulations, performed on a discrete lattice model, we include the relative timescale in the following way. We define a parameter $\omega$ which denotes the average number of times the landscape moves in between two particle movements. $\omega = 1$ corresponds to the case when one particle movement is followed by one landscape movement, on average. In all earlier studies of the LH model, $\omega = 1$ was considered \cite{chakrabortylarge, static, dynamic, chakrabortyUniversality}. Here, we consider any arbitrary $\omega$ within the $0 < \omega < \infty$ range. For $\omega \ll 1$, we have a landscape that moves much slower compared to the particles, and for $\omega \gg 1$, the landscape moves more rapidly compared to particle movement. We numerically measure various different short-range correlations in the system as a function of $\omega$. Surprisingly, we find a critical value $\omega = \omega_c$ at which all correlations in the system vanish, and the steady state is simply given by equiprobable measure. We explain this surprising result from exact calculation generalizing a formalism introduced by Mahapatra {\sl et al.} for the lattice model \cite{lightheavy}. We show that for every discrete configuration, it is possible to identify a bunch of incoming and outgoing transitions whose rates exactly balance each other for $\omega = \omega_c$ . This condition, “bunchwise balance” \cite{lightheavy}, ensures that in the steady state, all configurations occur with equal probability. We derive an exact expression for $\omega_c$ in terms of the coupling parameters between the particle and landscape dynamics.

To gain further insights into the above striking result, we consider the coarse-grained level description of the discrete LH model using hydrodynamics and examine the implications of the relative timescale there. In particular, for the LH model, the dynamical rules are such that the local particle density and local height gradient of the landscape remain conserved. Therefore, one can start from a set of two coupled continuity equations for these two conserved quantities, and within linear hydrodynamic approximation, one can solve these equations by diagonalizing the current Jacobian matrix $A$ and constructing the normal modes \cite{ferrari2013coupled}. The stationary fluctuations of the conserved variables are captured in the covariance matrix or compressibility matrix $K$. For a wide class of coupled driven systems with multiple conserved quantities, it was shown in \cite{Currentsymmetries} that the product $AK$ of the Jacobian and compressibility matrix must be symmetric, which is the nonequilibrium analog of Onsager’s reciprocity relation. However, to explicitly construct the matrix elements of $A$ and $K$, an exact knowledge of the stationary measure is required. We make an assumption that there exists a special value of the relative timescale $\omega = \omega^\ast$ where the steady state of the system satisfies product measure and all correlations vanish. Under this assumption, we can explicitly construct the matrices $A$ and $K$. Our calculations show that the symmetry condition is satisfied by the product $AK$ if and only if $\omega^\ast$ is equal to $\omega_c$ . Thus, not only for the discrete lattice model, even at a coarse-grained level, we have been able to exactly calculate
the expression for the critical value of the relative timescale which supports the product measure stationary state.

\section{Model description} \label{sec:model}

\begin{figure}[H]
\centering
\includegraphics[scale=0.5]{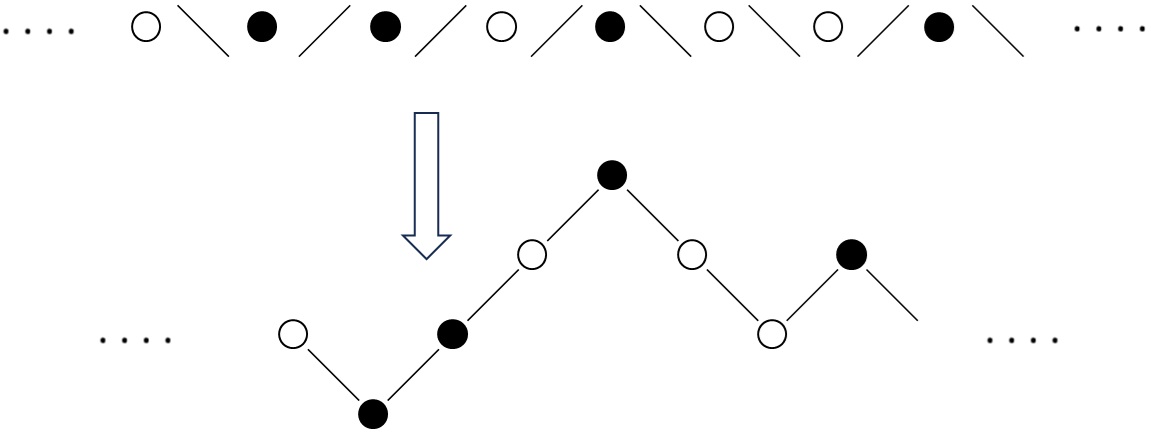}
\caption{A typical configuration in the LH model} \label{fig:hconf}
\end{figure}

The LH model, which is a short form for the light-heavy model, is defined on a lattice whose sites can be either occupied by an $H$ (heavy) particle or an $L$ (light) particle. The lattice bonds can have an orientation of $\pm \pi /4$. We denote the site occupancy by a variable $\eta_{j}$ , which takes the value $+1$ or  $-1$ if the site $j$ is occupied by an $H$ particle or an $L$ particle, respectively. We also use the filled circle \Large $\bullet$ \normalsize to represent an $H$-occupied site and empty circle \Large $\circ$ \normalsize to represent an $L$-occupied site. Similarly, the bond orientations are denoted by the variable $\tau_{j + \frac{1}{2}}$ which takes the value $-1$ if the bond between sites $j$ and $(j + 1)$ is an upslope bond, {\sl i.e.}, it has orientation $\pi/4$. For a downslope bond with orientation $-\pi/4$, the value of $\tau_{j + \frac{1}{2}}$ is $1$. The upslope and downslope bonds are also symbolically represented as $\slash$ and  $\backslash$, respectively. With each lattice site, we associate a “height” variable, defined as $h_i =- \sum_{j=1}^{i-1} \tau_{j + \frac{1}{2}}$ , as illustrated in Fig.\ref{fig:hconf}. From this figure, it is also clear that a local hill is represented by $\slash \backslash$ and a local valley by $\backslash \slash$.

\begin{figure}[h!]
\includegraphics[scale=0.4]{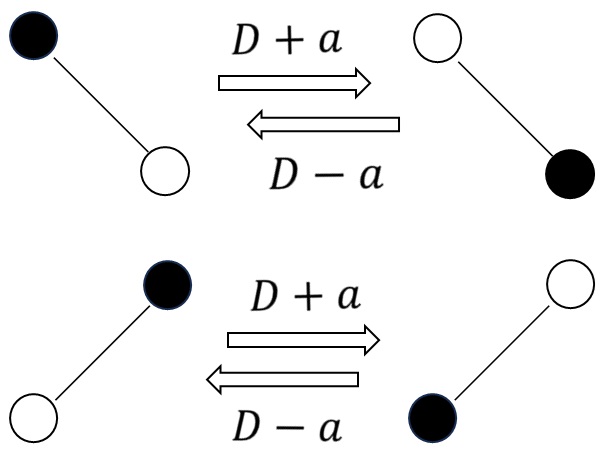}
\caption{Transition rules for the particles in the LH model} \label{fig:pmoves}
\end{figure}

The dynamics of the particles and the landscape are coupled in the following way. An $H$ particle can exchange position with an $L$ particle on its neighboring site, with the exchange rate depending on the slope of the intervening bond. $H$ particles tend to slide downward and $L$ particles show the opposite tendency. We define the particle sliding rates as follows:

\begin{equation} 
\label{eq:UpdateRule1}
\begin{split}
W(H\setminus L \rightarrow L\setminus H) & = D + a, \\
W(L\setminus H \rightarrow H\setminus L) & = D - a, \\
W(H\hspace{0.2em}/\hspace{0.2em}L \rightarrow L\hspace{0.2em}/\hspace{0.2em}H) & = D - a, \\
W(L\hspace{0.2em}/\hspace{0.2em}H \rightarrow H\hspace{0.2em}/\hspace{0.2em}L) & = D + a,
\end{split}
\end{equation}

with $0<a \le D$. In Fig. \ref{fig:pmoves}, we pictorially show these transitions. No other transitions are possible for the particles.
 
\begin{figure}
\includegraphics[scale=0.4]{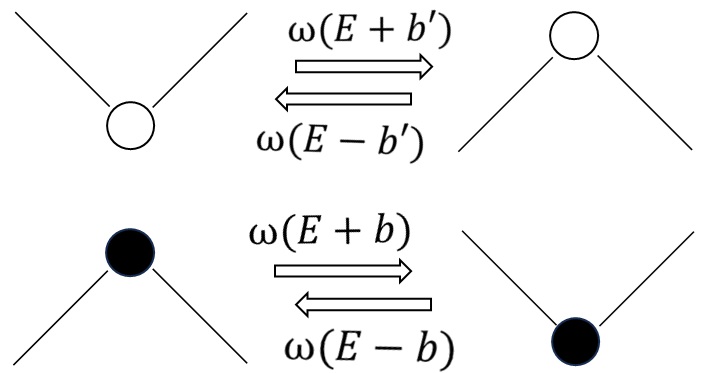}
\caption{Dynamical rules for the landscape movement in the LH model } \label{fig:smoves}
\end{figure}

The steady state particle current, defined as the rate of net flux of H particles across a bond, can be formally written as

\begin{equation}
\label{eq:particlej}
(D+a)[ P(H \backslash L)-P(L / H)] + (D-a)[P(H/L)-P(L \backslash H)],
\end{equation}

where $P(H \backslash L)$ denotes the probability of occurrence of the local configuration $H \backslash L$, etc. The landscape evolves by exchanging slopes between adjacent bonds, but the exchange rate depends on the occupancy of the site between those two bonds:

\begin{equation} 
\label{eq:UpdateRule2}
	\begin{split}
	W(/\hspace{0.1em}H\hspace{0.1em}\setminus \rightarrow \setminus \hspace{0.1em} H \hspace{0.1em}/) & = \omega(E + b), \\
	W(\setminus \hspace{0.1em} H \hspace{0.1em}/ \rightarrow / \hspace{0.1em}  H \hspace{0.1em}\setminus) & = \omega(E - b), \\
	W(/\hspace{0.2em}L\hspace{0.2em}\setminus \rightarrow \setminus \hspace{0.2em} L \hspace{0.2em}/) & = \omega(E - b'), \\
	W(\setminus \hspace{0.2em} L \hspace{0.2em}/ \rightarrow / \hspace{0.2em} L \hspace{0.2em} \setminus) & = \omega(E + b'),
	\end{split}
\end{equation} 

where the parameter $\omega$ controls the relative timescale between the landscape movement and the particle movement. We explain below how, in our simulations, $\omega$ introduces an update ratio of the particles and the landscape. Both the parameters $b$ and $b'$ in Eq. \ref{eq:UpdateRule2} are bounded between $-E$ to $E$. In Fig. \ref{fig:smoves} we pictorially show these transitions.

These landscape transitions result in a net upward or downward velocity of the landscape or, equivalently, a net rightward or leftward movement of the upslope bond through the system. The landscape current, defined as the rate at which an upslope bond moves through the system from left to right, is given by
 
\begin{equation}
\label{eq:slopej}
\omega (E+b) P(/ H \backslash) - \omega (E-b) P(\backslash H /) + \omega (E-b') P(/L \backslash) - \omega (E+b') P(\backslash L /). 
\end{equation}

From the dynamical rules shown in Eqs. \ref{eq:UpdateRule1} and \ref{eq:UpdateRule2}, it is clear that the local dynamics conserves the number of $H$ particles and $L$ particles, as well as the number of upslope and downslope bonds, in the system. We consider a lattice of size $N$ and a total number of $N_{up}$ upslope bonds and $(N-N_{up})$ downslope bonds. We denote the total number of $H$ particles as $N_H$; the number of $L$ particles then becomes $(N-N_H)$. We denote the density of $H$ particles as $\rho = N_{H}/N$ and the overall tilt of the landscape as $m = N_{up}/N$. We assume a periodic boundary condition on the lattice.

In all earlier studies of the LH model \cite{chakrabortylarge, static, dynamic, chakrabortyUniversality, lightheavy}, $\omega =1$ was assumed, but here we consider any arbitrary non-negative value of $\omega$. From Eq. \ref{eq:UpdateRule2}, it might appear that $\omega$ can be simply reabsorbed into the rate parameters and our present model then becomes identical to the previously considered version of the LH model \cite{chakrabortylarge, static, dynamic, chakrabortyUniversality, lightheavy}. But this is not the case. In particular, varying $\omega$ is not the same as varying $b,b'$ for fixed $E$, as was done previously \cite{chakrabortylarge, static, dynamic, chakrabortyUniversality}. In the rates $\omega (E \pm b)$ and $\omega (E \pm b')$, if we absorb $\omega$ into $E,b$ and $b'$ then as we vary $\omega$, it would mean varying all these three rescaled parameters $E,b$ and $b'$ proportionately. To ensure that rates are non-negative, rescaled $b$ and $b'$ should stay bounded between rescaled $[-E,E]$, {\sl i.e.} as $\omega$ is varied, the allowed range of variation of $b,b'$ must also change.

This is how we incorporate $\omega$ in our simulation algorithm. One Monte Carlo (MC) time-step consists of $N$ number of update trials. Before each update trial, we choose with probability $\omega/(1 + \omega)$ to carry out a landscape movement, and with probability  $1/(1 + \omega)$ to carry out a particle movement. If a landscape movement is decided, then the transition takes place with rate $(E \pm b)$ or $(E \pm b')$ depending on the local configuration (see Eq. \ref{eq:UpdateRule2}). If a particle movement is decided, then it is performed with rate $D \pm a$, as shown in Eq. \ref{eq:UpdateRule1}. Clearly, for large $\omega$, the landscape gets a chance to be updated much more frequently than the particles. For our choice of $D=E$, $\omega$  denotes average number of times the landscape moves in between two particle movements. $\omega \ll 1$ indicates a slowly moving landscape and $\omega \gg 1$ represents a fast moving landscape.

For $\omega =1$, some of the co-authors had constructed the phase diagram for the system in an earlier work \cite{chakrabortylarge}. For any $a >0$, the ordered and disordered phases were mapped out in the $bb'$ plane. It was found that for $(b+b') > 0$, the system is in various different kinds of ordered phases, where both particles and the landscape show long-range order. For $(b+b') < 0$, the system is in disordered phase, but short-range correlations still develop in the system. For $(b+b') = 0$, it follows from Eq. \ref{eq:UpdateRule2} that the landscape dynamics no longer depends on the particles, and the coupling becomes one way. This is a special limit and the system shows an unusual type of ordering for the particles here, where long-range order and strong fluctuations are simultaneously present \cite{chatterjee2006dynamics, fluctuationdominatedphase, Orderparameterscaling}.

\section{Simulation results on short-range correlations}

\begin{figure}[H]
\centering
\includegraphics[scale=0.8]{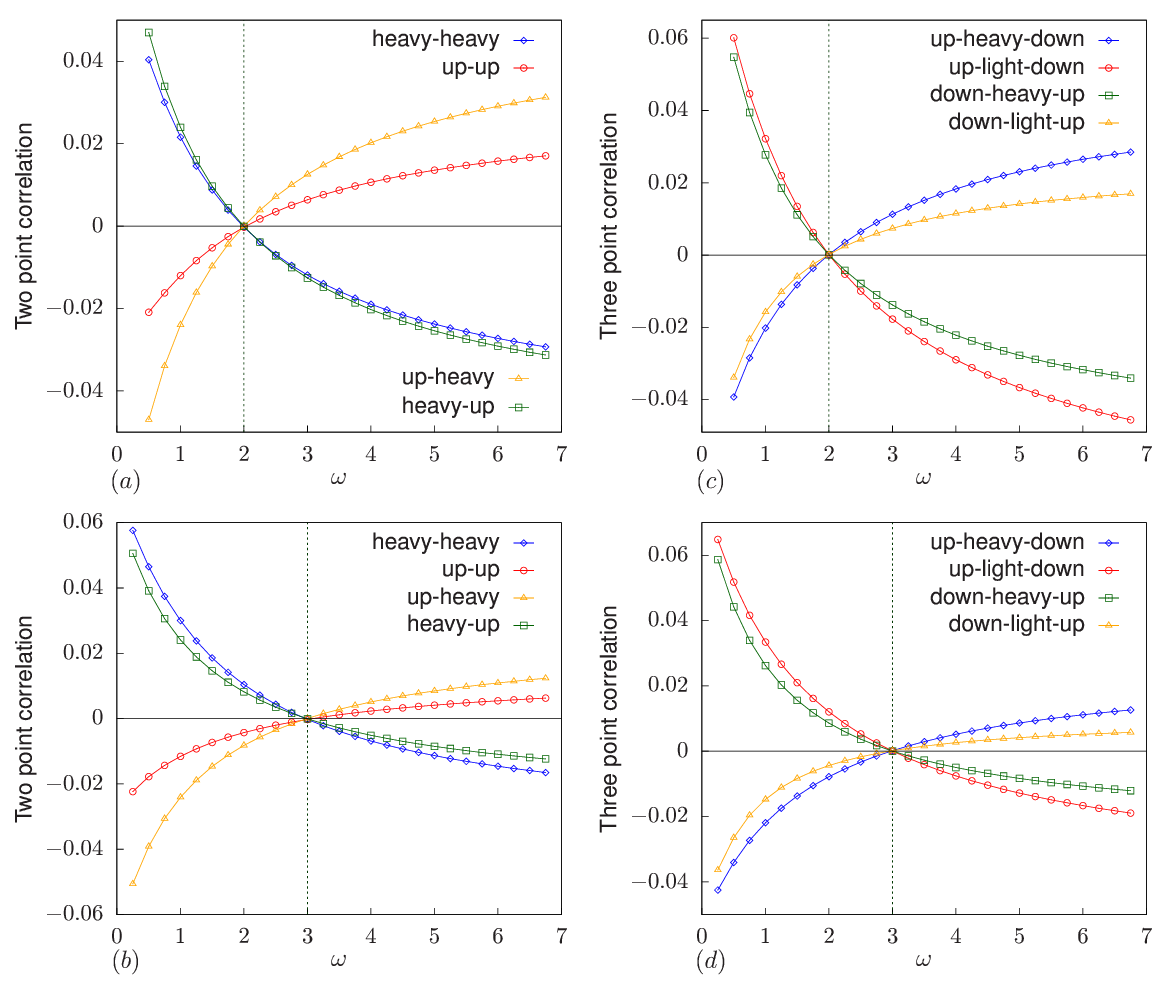}
\caption{Simulation results for two-point and three-point correlations. Top row corresponds to $b = -0.1$, $b' = -0.4$ and $a = 0.5$. Bottom row is for $b = 0.1$, $b' = -0.35$ and $a = 0.375$. For all panels, $N = 1024$, $N_H/N =\rho = 0.5$, $N_{up}/N=m = 0.5$.} \label{fig:Two_ThreePoint} 
\end{figure}

Starting from a random initial configuration, we perform $\sim N^{2}$ MC steps to reach the steady state and all our measurements are conducted in the steady state. We define the following nearest neighbor correlations. Consider $Prob(\eta_{i} = 1 ,\eta_{i + 1} = 1) $, which represents the joint probability that two consecutive lattice sites are occupied by $H$ particles. Similarly, $Prob (\tau_{i - \frac{1}{2}} = - 1 ,\tau_{i + \frac{1}{2}} = - 1)$ is the joint probability that there are two successive upslope bonds. In a similar way, one can define $Prob ( \eta_{i} = 1 , \tau_{i + \frac{1}{2}} = - 1)$ and $Prob(\tau_{i - \frac{1}{2}} = - 1 , \eta_{i} = 1)$. The two-point correlation functions can be defined from the joint probabilities. For example, the nearest neighbor correlation between two heavy particles is defined as  $Prob(\eta_{i} = 1 ,\eta_{i + 1} = 1) - (N_H/N)^2$. In Figs. \ref{fig:Two_ThreePoint}(a) and \ref{fig:Two_ThreePoint}(b), we plot the two-point correlations as a function of $\omega$. Surprisingly, we find that all four correlations vanish at a precise value of $\omega$. We denote this special point as $\omega_c$. For different choices of rate parameters $a,b,b'$, the value of $\omega_c$ is also different.

We also measure the variation of three-point correlations with $\omega$. While it is possible to define many different three-point correlations in the system, we present here data for only four different kinds, which appear in Eq. \ref{eq:UpdateRule2}. In Figs. \ref{fig:Two_ThreePoint}(c) and \ref{fig:Two_ThreePoint}(d) we present the data. Here, up-heavy-down refers to $Prob (\tau_{i - \frac{1}{2}} =  -1, \eta_{i} = 1,\tau_{i + \frac{1}{2}} = 1 ) - (\frac{N_{up}}{N})(1-\frac{N_{up}}{N})(\frac{N_H}{N})$ and other three correlations can also be similarly defined. We find that at exactly the same $\omega = \omega_c$ all three-point correlations vanish. We have verified it for a few other correlations too (data not shown here). To explain the above results, we generalize the formalism introduced in \cite{lightheavy} and show that at $\omega = \omega_c$ the system indeed satisfies equiprobable measure and no correlations exist in the system.

\section{Exact calculation for  $\omega_c$ for discrete lattice model} \label{sec:ExactCal}

In this section, we generalise the formalism introduced in \cite{lightheavy} to prove that at $\omega = \omega_c$, the system can be characterized by equiprobable measure in steady state. In other words, for a lattice of size $N$ with $N_H$ number of $H$ particles, $(N - N_H)$ number of $L$ particles, $N_{up}$ number of upslope bonds, and $(N - N_{up})$ number of downslope bonds, the total number of possible configurations is $\mathcal{N} = \binom{N}{N_H} \times \binom{N}{N_H}$ and for $\omega = \omega_c$, each configuration occurs with probability $1/\mathcal{N}$ in the steady state. Below, We exactly calculate an expression for $\omega_c$ in terms of the rate parameters $a,b,b'$.

The configurations in the LH model are discrete which are specified by the occupancy of the lattice sites and orientation or tilt of the lattice bonds. Let $p_c(t)$ is the probability to find the system in a particular configuration $c$ at time $t$. The time-evolution of $p_c(t)$ is governed by the master equation \cite{van1992stochastic}
\begin{equation} 
\frac{dp_c(t)}{dt} = \sum_{c' \neq c} \bigl[ r_{c' \rightarrow c} p_{c'}(t) - r_{c \rightarrow c'} p_{c}(t) \bigr].
\label{eq:master}
\end{equation}
Here, $r_{c' \rightarrow c}$ is the incoming transition rate from any other configuration $c'$ to $c$ and $r_{c \rightarrow c'}$ is the outgoing rate. In the steady state, the right-hand side of the above equation must vanish. Now, transition between two configurations are possible either via particle movement or via landscape movement. Particle movement involves exchange of positions between an $H$ and $L$ particle  at two neighboring sites. A landscape movement happens when a neighboring pair of upslope and downslope bonds exchange their orientations. Let us consider the quantity $(\eta_{j + 1} - \eta_{j})/2$ which can take the values $\pm 1 $ or $0$. For a local configuration of the form  $L / H$ or $L \backslash H$ the value is $+1$, for $H \backslash L$ or $H /L$ the value is $-1$ and for any other case where both neighboring sites are occupied by particles of the same species, the value is zero.  Clearly, no particle transition takes place if the value is zero. 

\subsection{Mapping of configuration to sequence of brackets and dots }

At this stage, we introduce an alternative way to represent a configuration, following the method introduced in \cite{lightheavy}. Instead of explicitly specifying the site occupancies, we use the symbol open parenthesis `$($' if $(\eta_{j + 1} - \eta_{j})/2$ is $+1$, a closed parenthesis `$)$' if it is $-1$, and a dot `$.$' if it is zero. Similarly, the quantity $(\tau_{j-\frac{1}{2}} - \tau_{j+\frac{1}{2}})/2 $ takes the value $+1$ for a local valley, $-1$ for a local hill, and zero for all other cases where no transitions are possible. We use an open angular bracket `$<$' for a local valley, a closed angular bracket `$>$' for a local hill and a dot `$.$' for other cases. In Table \ref{tab} we summarize the mapping.

\begin{table}[H]
\centering
\begin{tabular}{|c|c|c|c|c|}
\hline
\multicolumn{1}{|c|}{Variable} & Value & Local Configurations & Name & Symbol \\
\hline
\multirow{3}{*}{$\frac{1}{2}\bigl(\eta_{j + 1} - \eta_{j}\bigr)$} & 1 & $\circ / \bullet  \hspace{0.15cm} ; \hspace{0.15cm} \circ \setminus \bullet$ & open parenthesis & ( \, \\
& 0 & $\circ / \circ \hspace{0.15cm} ; \hspace{0.15cm} \circ \setminus \circ \hspace{0.15cm} ; \hspace{0.15cm} \bullet \setminus \bullet \hspace{0.15cm} ; \hspace{0.15cm} \bullet / \bullet$ & dot & $\cdot$ \\
& -1 & $\bullet / \circ \hspace{0.15cm} ; \hspace{0.15cm} \bullet \setminus \circ$ & closed parenthesis & )\, \\
\hline
%\multicolumn{1}{|c|}{Merged 2} & Column 2 & Column 3 & Column 4 & Column 5 \\
%\hline
\multirow{3}{*}{$\frac{1}{2} \bigl(\tau_{j-\frac{1}{2}} - \tau_{j+\frac{1}{2}}\bigr)$} & 1 & $\backslash \circ \slash \hspace{0.15cm} ; \hspace{0.15cm} \backslash \bullet \slash$ & open angular bracket & $\langle$ \\
& 0 & $\slash \bullet \slash \hspace{0.15cm} ; \hspace{0.15cm} \slash \circ \slash \hspace{0.15cm} ; \hspace{0.15cm} \backslash \bullet \backslash \hspace{0.15cm} ; \hspace{0.15cm} \backslash \circ \backslash$ & dot & $\cdot$ \\
& -1 & $\slash \circ \backslash \hspace{0.15cm} ; \hspace{0.15cm} \slash \bullet \backslash$ & closed angular bracket & $\rangle$ \\
\hline
\end{tabular}
\caption{Representation of configuration in the LH model as a sequence of brackets and dots} \label{tab}
\end{table}

\begin{figure}[H]
\centering
\includegraphics[scale=0.45]{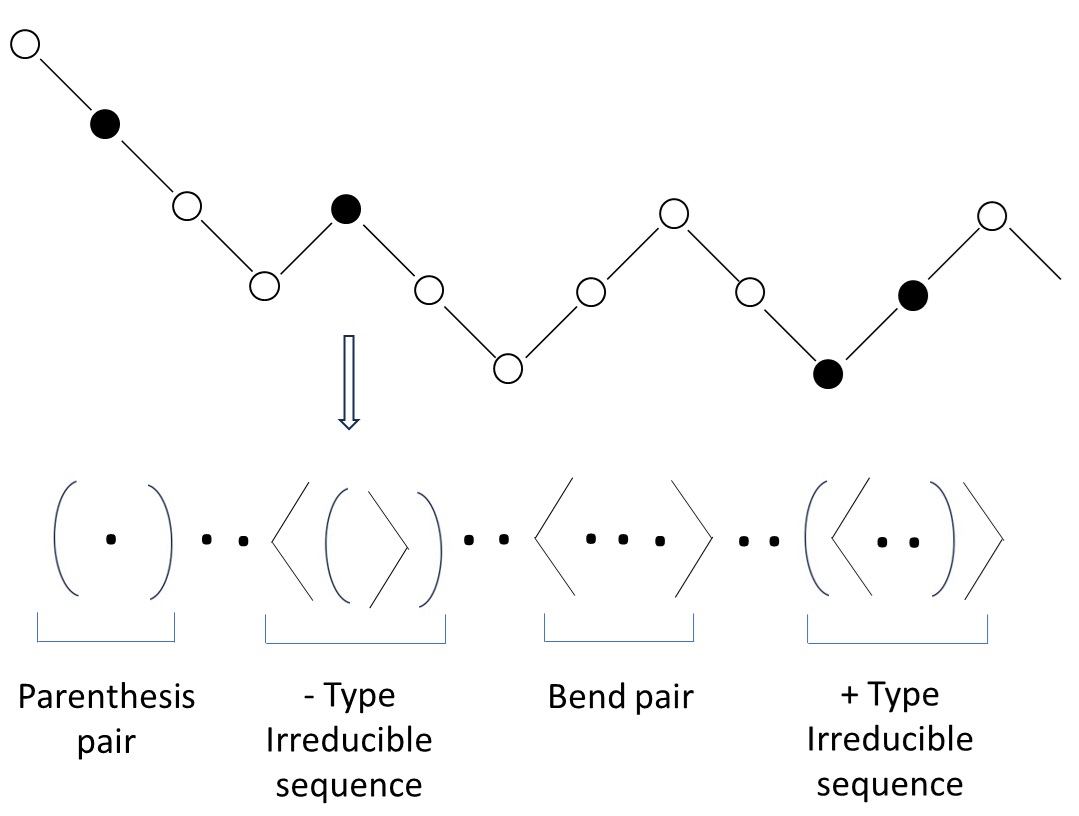}
\caption{Description of a configuration in terms of parentheses and bends and different type of segments arising in it.}   \label{fig:map}
\end{figure}

Using the above method, one can map any configuration in the LH model as a unique sequence of parentheses, angular brackets and dots. In Fig. \ref{fig:map} we illustrate this process. The sequence follows some constraints though. Two open (closed) parentheses can not occur in succession, there must be a closed (open) parenthesis between them. Similar restriction holds for angular brackets too. Due to periodic boundary condition, each configuration must have equal number of open and closed parentheses, and an equal number of open and closed angular brackets. However, the total number of parentheses may or may not be equal to the total number of angular brackets in a configuration. Moreover, the dynamics does not conserve these total numbers and they can change with time, even though $N_H$ and $N_{up}$ remain conserved. 

\subsection{Grouping brackets and dots into segments }

In the steady state, the total incoming flux to any configuration $c$ must be balanced by the total outgoing flux from $c$. Here we classify all incoming and outgoing transitions for a general configuration $c$ into groups. For this, we follow the protocol introduced in \cite{lightheavy}. We start from any random point in the sequence of parentheses, angular brackets and dots which represents configuration $c$, and move rightward through the sequence. Every time we encounter an open (closed) parenthesis or open (closed) angular bracket, we move towards right (left), and keep moving in that direction until we find a closed (open) bracket of the same type. While doing so, if we encounter other type of open (closed) unpaired bracket, we move in rightward direction (leftward direction) until we get a closed (open) bracket of that type. This procedure is continued until a segment of minimum size is identified, where corresponding to each open bracket, there is a closed bracket of the same type. After a segment has been completed, we start a new segment from the current position and repeat the process. This method divides a sequence into a few non-overlapping segments and also ensures that this segmentation is unique since the protocol ensures that every bracket belongs to the smallest group of open and closed brackets \cite{lightheavy}. In Fig. \ref{fig:map} we illustrate this method. Next, we show how within each segment the steady state flux balance condition is satisfied.

\subsection{Pairwise balance}

\begin{figure}[H]
\centering
\includegraphics[scale = 0.6]{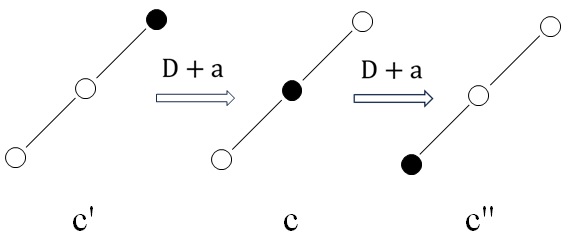} 
\caption{Pairwise balance for a sequence enclosed by a pair of parentheses. For every incoming transition, there is an outgoing transition of the same rate.} \label{fig:pairwise}
\end{figure}

The simplest possible segment contains only one pair of open and closed brackets of the same type, {\sl i.e.} either $(...)$ or $<....>$. Let us first consider $(...)$ which represents a local configuration $c$ where only particle movements are  possible, and no landscape movement is allowed. As we illustrate in Fig. \ref{fig:pairwise} for every such configuration $c$, one can identify a unique pair of configurations $c'$ and $c''$, such that $r_{c' \to c} = r_{c \to c''}$. Absence of any angular brackets rules out the existence of local hills and valleys in $c$. For example, if we assume all lattice bonds are downhill, then $c$ can be entered (exited) by an upward (downward) slide $H$ particle at the left (right) end. Thus for every incoming transition, there exists an outgoing transition of the very same type. This condition is known as the `pairwise balance' \cite{schutz1996pairwise, lightheavy}. More precisely, if $r_{c' \to c} = r_{c \to c''}$ is satisfied, then $p_{c'} = p_c$ indeed makes the right hand side of Eq. \ref{eq:master} zero for any values of $a,b,b'$. In a similar way, one can show that pairwise balance with equiprobable measure condition is satisfied for all $<...>$ segments.

\begin{figure}[H]
\centering
\includegraphics[scale = 0.45]{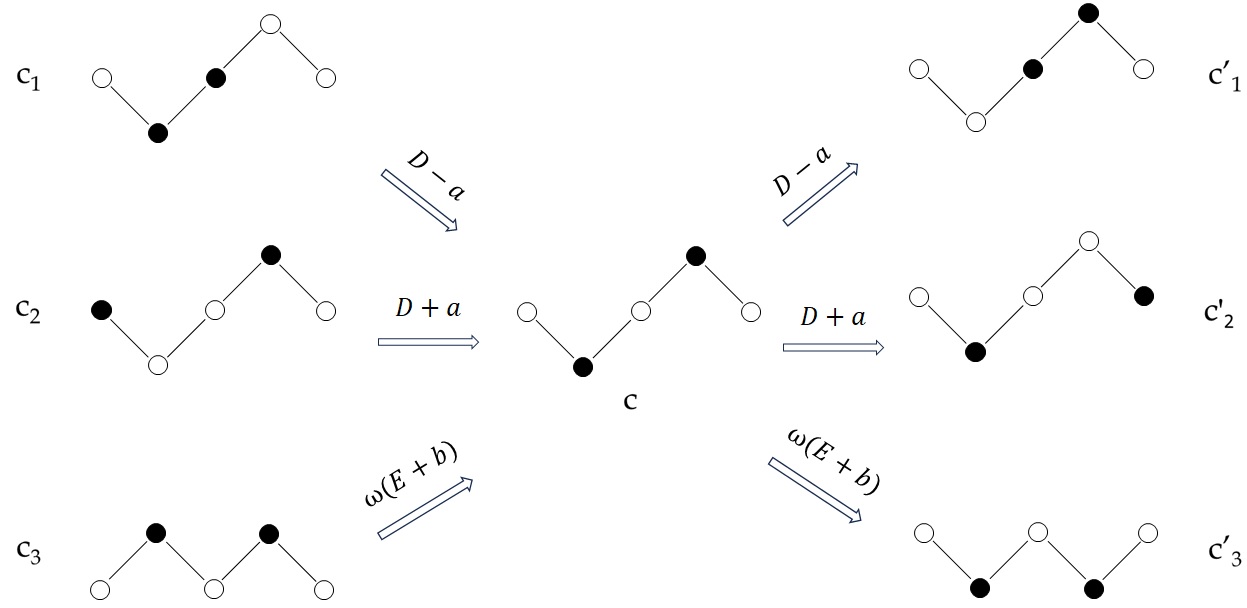}
\caption{Pairwise balance for a local configuration $c$ represented by $ (\,..\langle..)\,..(\,..\rangle..)\,$} \label{fig:ComplicatedPair}
\end{figure}

Even for a more general type of segment, where both parentheses and angular brackets are present, pairwise balance condition can be satisfied as long as the leftmost opening bracket and the rightmost closing bracket are of the same type. To show this we use the method of reduction introduced in \cite{lightheavy} where a given segment can be shortened by eliminating those pairs of brackets which represent transitions that satisfy pairwise balance. For example, consider the segment $(..<..)..(..>..)$. It can be easily shown that the pair $)..($, appearing in the middle, satisfies pairwise balance. The proof runs along exactly the same lines as illustrated in the previous paragraph for $(..)$ segment. After eliminating this pair, the segment is reduced to $(..<..>..)$ and after eliminating the pair of angular brackets $<...>$ which obviously satisfy pairwise balance, we are left with $(..)$, another pairwise balanced segment. In Fig. \ref{fig:ComplicatedPair} we explicitly show, without using the method of reduction, that indeed the local configuration represented by the above sequence segment satisfies pairwise balance. In a similar way, one can show that pairwise balance is satisfied for the segment $<..(..>..<..)..>$.

\subsection{Bunchwise balance}

\begin{figure}[H]
\centering
\includegraphics[scale=0.6]{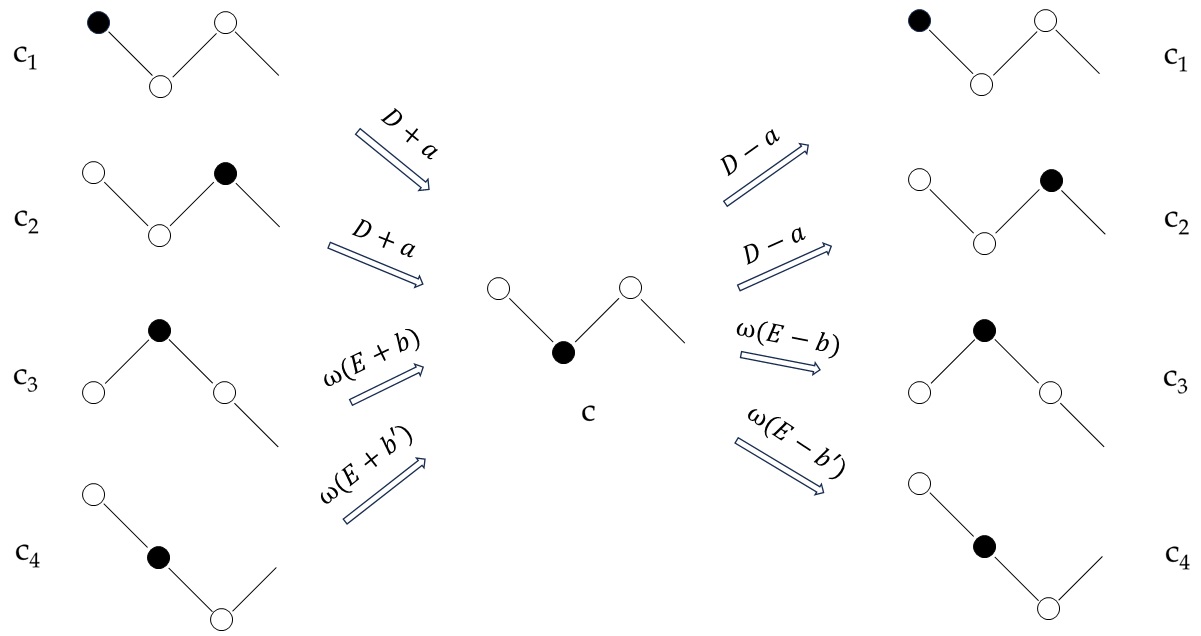}
\caption{Incoming and outgoing transitions for a plus type irreducible sequence. }   \label{fig:tranptype}
\end{figure}

However, if a segment has leftmost opening bracket and rightmost closing bracket of opposite types, then it can not be reduced completely. After eliminating all pairwise balanced brackets like $<..>$, $>..<$, $(..)$ and $)..($ which may have been present in the middle of the segment, we are left with either $(..<..)..>$ or $<..(...>..)$, which can not be reduced any further. These are known as irreducible sequence of $+$ type and $-$ type, respectively. They do not satisfy pairwise balance \cite{lightheavy}. However, both these irreducible sequences represent local configurations that can be reached in four different ways, two of which involve particle exchange and the other two involve landscape movement. In Fig. \ref{fig:tranptype} we explicitly show this for $+$ type irreducible sequence. The particle transitions are always of the same type, and if one landscape transition consists of heavy (light) hill flipping to a heavy (light) valley, then the other one corresponds to light (heavy) valley flipping to a light (heavy) hill. The configurations can be exited by four transitions, which are exactly the reverse transitions of the incoming ones. If equiprobable measure has to be satisfied in steady state, then the sum of all incoming transition rates must be equal to the sum of all outgoing transition rates. This leads to the condition of `bunchwise balance'
\begin{equation}
2(D+a) + \omega (E+b) + \omega (E+b') = 2(D-a) + \omega (E-b) + \omega (E-b')  
\end{equation}
which gives 
\begin{equation}
\omega = \omega_c= -\frac{2a}{b+b'}. \label{eq:wc}
\end{equation}
For $-$ type irreducible sequence also, condition of bunchwise balance gives the exactly same expression for the critical time scale. In the special case of $\omega_{c} = 1$, the bunchwise balance condition simplifies further. In this case, the total incoming (outgoing) transition rate through particle exchange becomes equal to the total outgoing (incoming) transition rate through slope update, such that $2(D+a) = (E-b) + (E-b')$ which gives $2a + b + b' = 0$, the condition derived in \cite{lightheavy}. 

\begin{figure}[H]
\centering
\includegraphics[scale=1.0]{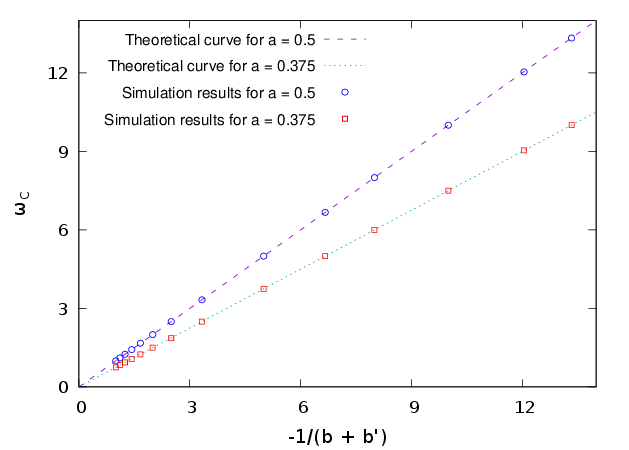}
\caption{Comparison between theory and simulations. Discrete points show $\omega_{c}$ measured from simulations and lines show exact result from Eq. \ref{eq:wc}. Here, $\rho=m=0.5$ are used.  } \label{fig:Matching}
\end{figure}

Eq. \ref{eq:wc} shows that it is only in the disordered phase, {\sl i.e.} for $(b+b') < 0$ one has a positive, finite value of $\omega_c$. As one approaches the $(b+b') =0$ line, the value of $\omega_c$ becomes infinitely large. This is expected, since the particles show long range order here and product measure cannot be satisfied. For $(b+b') > 0$ both particles and the landscape show long range order and $\omega_c$ takes unphysical negative values here. In Fig. \ref{fig:Matching} we compare our exact result with our simulation observation and find excellent agreement.

\section{Relative time-scale in hydrodynamics}
\label{sec:hydro}
In the previous section, an exact expression for $\omega_c$ was derived for a discrete lattice model using the condition of flux balance in steady state. In this section, we present an alternative derivation in the continuum limit where we coarse-grain the LH model and use hydodynamic expansion. As discussed in Sec. \ref{sec:model} the dynamical rules are such that the number of $H$ particles and the number of upslope bonds are locally conserved. We denote the local density of $H$ particles as $\rho(x,t)$ and that of upslope bonds as $m(x,t)$. As the particles move around and the landscape fluctuates, these densities change and their time-evolution is described by the following continuity equations
\begin{eqnarray}
\frac{\partial \rho}{\partial t} &=& -\frac{\partial J_\rho (\rho, m) }{\partial x}  \\
\frac{\partial m}{\partial t} &=& -\frac{\partial J_m (\rho, m) }{\partial x}
\end{eqnarray}
where $J_\rho$ and $J_m$ are local currents of $H$ particles and upslope bonds, respectively. Due to coupled dynamics, these currents depend on both $\rho$ and $m$. Here, we have used the assumption of local equilibrium, {\sl i.e.} local currents depend on $x,t$  only through their dependence on $\rho, m$ and do not have any explicit space-time dependence. Then the continuity equations can be written as
\begin{eqnarray}
\frac{\partial \rho}{\partial t} + \frac{\partial J_\rho}{\partial \rho} \frac{\partial \rho }{\partial x} +  \frac{\partial J_\rho}{\partial m} \frac{\partial m }{\partial x} =0 \\
\frac{\partial m}{\partial t} + \frac{\partial J_m}{\partial \rho} \frac{\partial m }{\partial x} +  \frac{\partial J_m}{\partial m} \frac{\partial m }{\partial x} =0.
\end{eqnarray}
By defining a two dimensional vector $\vec{\psi}(x,t)$, whose components are $\rho(x,t)$ and $m(x,t)$, the above equations can be written in a more compact form as
\begin{equation}
\frac{\partial \vec{\psi}}{\partial t} + {A}\frac{\partial \vec{\psi}}{\partial x} = 0
\end{equation}
where $A$ is the current Jacobian defined as
\begin{equation}
{A} = \begin{pmatrix}
\frac{\partial J_\rho}{\partial \rho} & \frac{\partial J_\rho}{\partial m} \\
\frac{\partial J_m}{\partial \rho} & \frac{\partial J_m}{\partial m}
\end{pmatrix}.
\end{equation}
Expanding components of $\vec{\psi}(x,t)$ around their conserved global values $\psi_\alpha (x,t) = \psi_\alpha ^0 + u_\alpha (x,t)$ and retaining only linear terms in $u_\alpha$ we get
\begin{equation}
\frac{\partial \vec{u}}{\partial t} + {\bm A}^0 \frac{\partial \vec{u}}{\partial x} = 0
\end{equation}
where ${A}^0$ is a $2 \times 2$ matrix whose elements are those of $\bm A$ evaluated at $\rho_0=N_H/N$ and $m_0 = N_{up}/N$. To solve this equation, we diagonalize ${A}^0$ and obtain the normal modes. However, this requires explicit knowledge of $J_\rho$ and $J_m$ as a function of $\rho$ and $m$. From Eqs. \ref{eq:particlej} and \ref{eq:slopej} in Sec. \ref{sec:model} it follows that the exact expressions for certain short ranged correlations are needed, which is generally hard to obtain. At this stage, we make an assumption that there exists a specific $\omega = \omega^\ast$ for which product measure holds and these correlations factorize. Then from Eqs. \ref{eq:particlej} and \ref{eq:slopej} we can write $J_\rho$ and $J_m$ as (scaled by a factor of ($1+\omega^\ast$))
\begin{equation} \label{eq:jw}
	\begin{split}
	J_{\rho} & =  2a\rho(1 - \rho)(1 - 2m) \\
	J_{m} & =  \omega^\ast 2 m(1 - m)\bigl[\rho(b + b') - b'\bigr].
	\end{split}
	\end{equation}
	The current jacobian matrix $A^0_{\alpha\beta}$ = $\dfrac{\partial J_\alpha}{\partial \rho_\beta}$ then has the form
	\begin{equation} 
	\label{eq:Jacobian}
	\centering
	{A^0} =  \begin{bmatrix}
	2a(1 - 2m_{0})(1 - 2\rho_{0}) & -4a\rho_{0}(1-\rho_{0})\vspace{0.15cm}\\
	2\omega^\ast m_{0}(1-m_{0})(b+b') & \omega^\ast (1-2m_{0})\bigl(2\rho_{0}(b+b') - 2b'\bigr)
	\end{bmatrix}.
	\end{equation}
which has real, distinct eigenvalues. Therefore, we have a system of strictly hyperbolic conservation laws. For such systems it was shown in \cite{Currentsymmetries} using time-reversal invariance that the product of current Jacobian matrix and the compressibility matrix must be symmetric, 
\begin{equation}
\label{eq:Onsagar}
	{A}^0 K = K({A}^0)^T
\end{equation}
where $({A}^0)^T$ is transpose of ${A}^0$.

Now in a canonical ensemble which we use in the simulations, the global conservation of $N_H$ and $N_{up}$ does give rise to correlations in a finite size system, even when equiprobable measure is satisfied in steady state. Product measure condition, which we used in writing Eq. \ref{eq:jw}, actually holds for a grand canonical ensemble with a fluctuating  $N_H$ and $N_{up}$. These fluctuations are characterized by a symmetric covariance matrix (also known as compresssibility matrix) $K$ whose elements are \cite{fibonacci, popkovuniversality, popkovexact}
\begin{equation}
K = \frac{1}{N} \begin{pmatrix} 
\langle (N_H - \rho N)^2 \rangle & \langle (N_H - \rho N)(N_{up}-mN)\rangle \\
\langle (N_{up}-mN)(N_H - \rho N)\rangle & \langle (N_{up}-mN)^2\rangle \end{pmatrix}
\end{equation}
In the case when product measure is satisfied, $K$ becomes
\begin{equation}
\label{eq:Comp}
K = \begin{pmatrix}
\rho_{0}(1-\rho_{0}) & 0 \\
0 &  m_{0}(1-m_{0}). 
\end{pmatrix}
\end{equation}

Substituting forms of ${A^0}$ and $K$ in Eq.\ref{eq:Onsagar} we get
\begin{equation} 
	\label{eq:Onsagar2}
	\centering
	{A}^0 K - K({A}^0)^T =  \begin{bmatrix}
	0 & -2\rho_{0}(1-\rho_{0})m_0(1-m_0)\bigl(2a + \omega^\ast(b+b')\bigr)\vspace{0.15cm}\\
	2\rho_{0}(1-\rho_{0})m_0(1-m_0)\bigl(2a + \omega^\ast(b+b')\bigr) & 0
	\end{bmatrix}
	\end{equation}
The right  hand side of Eq. \ref{eq:Onsagar2} vanishes only when $\omega^\ast = -\dfrac{2a}{(b+b')}$,  which matches exactly with $\omega_c$ derived in sec.\ref{sec:ExactCal}.

\section{Conclusions}

To gain insights into the behavior of coupled driven systems, one useful tool is to make one system evolve faster or slower than the other, by introducing a relative timescale in their dynamics. This method has been used in quite a few recent studies \cite{sadhu2016actin, sadhu2018actin, passiveslider1, passiveslider2, singleactiveslider} and various interesting effects of the relative timescale were observed. In \cite{passiveslider1, passiveslider2}, noninteracting particles sliding on a fluctuating landscape were considered in the limit of one-way coupling, where the landscape dynamics remains unaffected by the particles, and it was found that depending on whether the particles move faster or slower compared to the landscape, the scaling description of several
steady state correlations changes. In \cite{singleactiveslider}, a single particle was considered on a fluctuating energy landscape in the presence of a two-way coupling, and qualitatively different long-time behavior were observed, depending on the landscape to slider timescale ratio.
In this work, we have examined the effect of introducing a relative timescale $\omega$ in a coupled time evolution of hard-core particles on a fluctuating landscape. We find a critical value $\omega_c$ of the relative timescale at which all configurations are equally likely in the steady state. We prove this result exactly for a discrete lattice model, as well as for a coarse-grained model. Our proof for the case of a discrete lattice model relies on generalizing a formalism introduced in \cite{lightheavy}, according to which, we break up each configuration into nonoverlapping segments and identify the incoming and outgoing transitions for each segment. For certain segments, we find that for each incoming transition, there exists an outgoing transition of exactly the same type. This condition is known as pairwise balance \cite{schutz1996pairwise} and it follows from Eq. \ref{eq:master} that the equiprobable measure is satisfied in the steady state. However, there are some special types of segments where the incoming and outgoing transitions are exactly the reverse of each other. For these cases, the equiprobable measure can be satisfied if and only if the total incoming rate and outgoing rate are equated, which yields the $\omega = \omega_c$ criterion. For $\omega \neq \omega_c$, the equiprobable measure does not hold and local correlations are nonzero in general. It was not possible to calculate these correlations exactly, but their qualitative behavior can be explained by following the line of argument presented in \cite{lightheavy}. In Appendixes \ref{app:corr} and \ref{app:clust}, we have included detailed discussions of this. Our study opens up the intriguing possibility to control the stationary measure of a coupled driven system by tuning its relative timescale. This idea can be implemented for other model systems as well. In particular, our calculation in the coarse-grained model is quite generic and, for any other system which allows hydrodynamic expansion, similar calculations can readily be performed. It will be of interest to see if such calculations yield a critical timescale for other systems. Finally, for systems with $n > 2$ conserved modes, there can be $(n - 1)$ relative timescales. How their interplay affects the correlation functions and whether it allows for a product measure steady state are interesting questions and offer a promising direction for future research.

\section{Acknowledgments}

We acknowledge useful discussions with Gunter M. Schütz. C.K. acknowledges a research fellowship (Grant No. 09/0575(12571)/2021-EMR-I) from the Council of Scientific and Industrial Research (CSIR), India.

\appendix
\section{Variation of correlations with \large $\omega$ \normalsize} 
\label{app:corr}

As shown in Fig.\ref{fig:Two_ThreePoint}, different local short-ranged correlations show different natures of variation with $\omega$. In case of three point correlations, $\slash$\Large$\circ$\normalsize$\backslash$ and $\backslash$\Large$\bullet$\normalsize$\slash$ start from a positive value, decrease with $\omega$, pass through zero at $\omega = \omega_c$ and become negative for larger $\omega$. The other three point correlations, $\slash$\Large$\bullet$\normalsize$\backslash$ and $\backslash$\Large$\circ$\normalsize$\slash$ show exactly the opposite trend. They are negative for $\omega < \omega_c$ and positive for $\omega > \omega_c$. These variations can be easily explained by considering how the bunchwise balance of incoming and outgoing transition rates is violated when $\omega \neq \omega_c$. As shown in Fig.\ref{fig:tranptype}, the sum total of all incoming transition rates for a $+$ type irreducible sequence is  $2(D+a)+\omega (2E+b+b')$, which is greater than the sum of all outgoing transition rates $2(D-a)+\omega (2E-b-b')$, if $\omega < \omega_c$. For a $-$ type sequence, similarly, one can show that the total outgoing rate is larger than the incoming ones. Therefore, for $\omega < \omega_c$ the steady state probability to find a $+$ type sequence is more than $-$ type. As a result, those correlations which occur in $+$ type sequences are positive for  $\omega < \omega_c$ and negative for $\omega > \omega_c$, while those correlations which are absent from $+$ type sequences but can be found in $-$ type sequences, show the opposite trend.

\begin{figure}[H]
\begin{center}
\includegraphics[scale = 1.0]{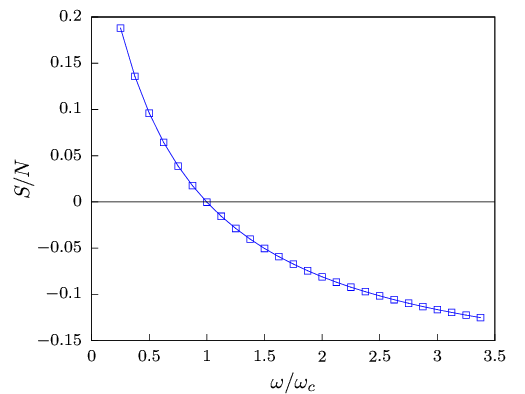}
\caption{Variation of $S$ with relative time-scale for $b = -0.1, b' = -0.4, a = 0.5$, $\rho_0=m_0=0.5$ and $N = 1024$. }
\label{fig:cross}
\end{center}
\end{figure}

At $\omega = \omega_c$ the $+$ type and $-$ type irreducible sequences are equally probable. The following quantity, known as cross-correlation function captures this clearly \cite{lightheavy}. This is defined as
\begin{equation}
{\mathcal{S}}(\{\eta,\tau\})=\sum_{j=1}^N \frac{1}{2}(\eta_{j+1}-\eta_j)\tau_{j+\frac{1}{2}} 
\label{eq:triad}
\end{equation} 
which depends on the specific configuration $\{\eta,\tau\}$. The cross-correlation function can also be defined as ${\mathcal{S'}}(\{\eta,\tau\}) = \sum_{j=1}^{N} \frac{1}{2} \bigl (\tau_{j - \frac{1}{2}} - \tau_{j + \frac{1}{2}} \bigr )\eta_{j}$ which has the same value as ${\mathcal{S}}(\{\eta,\tau\})$ for any configuration with periodic boundary conditions. ${\mathcal{S}}$ and ${\mathcal{S'}}$ give a value of $0$ when we calculate them only over a parenthesis pair or a bend pair. A segment that satisfies pairwise balance has a vanishing contribution towards the sum in Eq. \ref{eq:triad}. The only nonzero contribution comes from irreducible sequences of $+$ and $-$ type, which contribute $2$ and $-2$, respectively. If $N_+ (\{\eta,\tau\}) $ and $N_- (\{\eta,\tau\})$ denote the number of $+$ type and $-$ type irreducible sequences in a configuration, then
\begin{equation}  
{\mathcal{S}}(\{\eta,\tau\}) = 2[N_+(\{\eta,\tau\})-N_-(\{\eta,\tau\})].
\end{equation}
When averaged over steady state ensemble, we get $S = \langle {\mathcal{S}}(\{\eta,\tau\}) \rangle = 2 (\langle N_+ \rangle-\langle N_- \rangle)$, where $\langle N_\pm \rangle$ are the average number of $+$ type and $-$ type irreducible sequences in steady state. In Fig. \ref{fig:cross} we plot $S$ (scaled by system size) as a function of the relative time-scale and find that for $\omega < \omega_c$ when $+$ type sequences are more probable, $S$ is positive, while for $\omega > \omega_c$, when $-$ type sequences are more probable, $S$ is negative. This is consistent with the above argument.

\section{height fluctuations and cluster size distributions }
\label{app:clust}

\begin{figure}[H]
		\centering
		\includegraphics[scale=0.8]{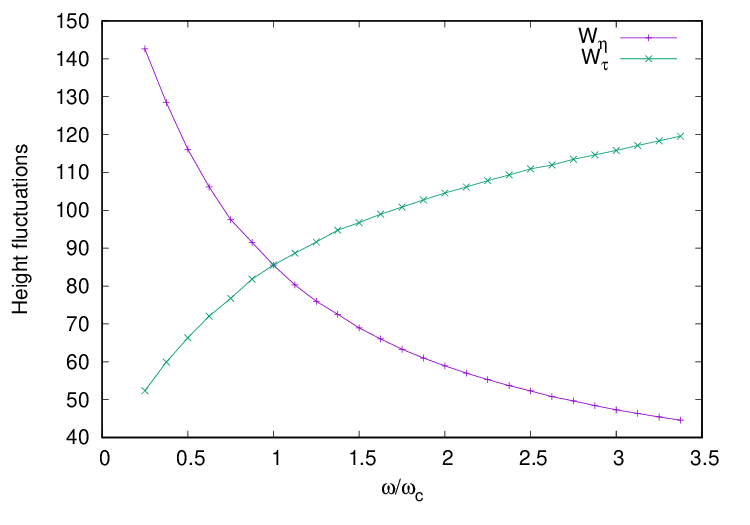}
		\caption{Variation of $W_\eta$ and $W_\tau$ with $\omega$ for for $b = -0.1, b' = -0.4$ and $a = 0.5$ with $\rho_0 = m_0 = 0.5$. }
		\label{fig:width}
	\end{figure}

\begin{figure}[H]
		\centering
		\includegraphics[scale = 0.8]{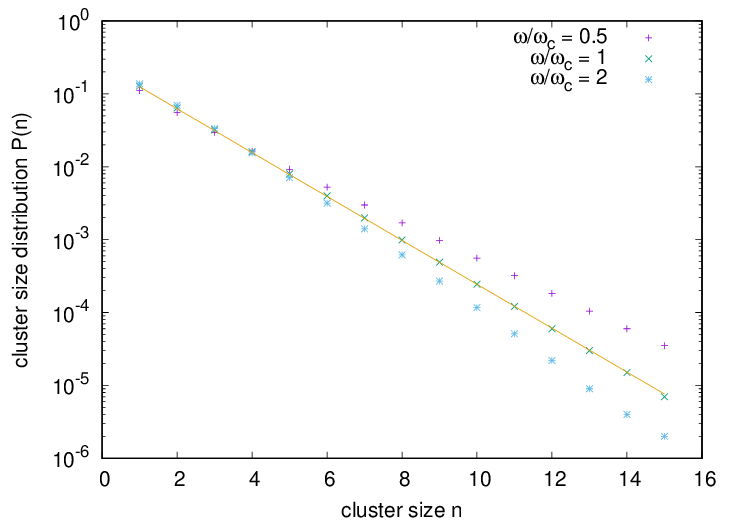}
		\caption{Cluster size distribution of particles at different $\omega$ for $b = -0.1, b' = -0.4$ and $a = 0.5$ with $\rho_0 = m_0 = 0.5$. The solid line represents $P(n)$ for product measure.}
		\label{fig:clust}
	\end{figure}

To further probe the effect of relative time-scale on the steady state of the system, we define two `height' variables for site occupancy and bond orientation as follows.  
\begin{equation}
h_\eta (i) = \sum_{j=1}^i \eta_j \;\;\;\;\;\; {\text{and}} \;\;\;\;\;\; h_\tau (i)= -\sum_{j=1}^i \tau_{j+1/2}.
\end{equation}
Let ${ W}_\eta$ and ${ W}_\tau$ be the width of the corresponding height profiles, defined as 
\begin{equation}
		{W}_{\eta} = \biggl[ \frac{1}{L} \Bigl \langle \sum_{i=1}^{L} \bigl[h_{\eta}(i) - \overline{h_{\eta}}\hspace{0.1cm}\bigr]^{2} \Bigr \rangle\biggr]^{1/2} \hspace{0.2cm} ; \hspace{0.2cm} {W}_{\tau} = \biggl[ \frac{1}{L} \Bigl \langle \sum_{i=1}^{L} \bigl[h_{\tau}(i) - \overline{h_{\tau}}\hspace{0.1cm}\bigr]^{2} \Bigr \rangle\biggr]^{1/2}
	\end{equation}
where the overhead bar denotes averaging over all lattice points in a given configuration and the angular brackets denote averaging over different steady state configurations. In Fig. \ref{fig:width} we plot ${W}_{\eta}$ and ${W}_{\tau}$ for different $\omega$ values and find two opposite trends. While ${W}_{\eta}$ decreases with $\omega$ (purple points), ${W}_{\tau}$ increases (green points) and they intersect at $\omega = \omega_c$. To explain this behavior, we note that the tendency of $H$ particles to slide down to the local valleys is exactly balanced by their tendency to destabilize the valleys by landscape movement at $\omega = \omega_c$. This gives rise to equiprobable measure. For $\omega$ smaller than $\omega_c$, the landscape movement is slower and the $H$ particles get a chance to form local clusters around the valleys. Although this heterogeneity in particle occupancy does not give rise to long range order in the system, the width ${W}_{\eta}$ is still larger than its product measure value. To verify this argument, we explicitly measure the cluster size distribution of $H$ particles (Fig. \ref{fig:clust} ) and show that for $\omega < \omega_c$, medium and large size clusters are more probable than the product measure case. Simiarly, for $\omega > \omega_c$ the landscape moves faster and heavy valleys are quickly destabilized. This makes it more difficult for the $H$ particle to clump together  and isolated $H$ particles are found more frequently now. This brings down  ${W}_{\eta}$. In a similar way, variation of 
${W}_{\tau} $ can also be explained. 

%\bibliographystyle{unsrt}
%\bibliography{review3} 

\end{document}